\documentclass[12pt]{iopart}

\usepackage{graphicx}

\begin{document}

\title{Advanced tools for smartphone-based experiments: phyphox}

\author{S Staacks$^1$, S H\"utz$^1$, H Heinke$^1$, C Stampfer$^1$}
\address{$ˆ1$ Institute of Physics I and II, RWTH Aachen University,
52062 Aachen, Germany}
\ead{staacks@physik.rwth-aachen.de}

\begin{abstract}
The sensors in modern smartphones are a promising and cost-effective tool for experimentation in physics education, but many experiments face practical problems. Often the phone is inaccessible during the experiment and the data usually needs to be analyzed subsequently on a computer. We address both problems by introducing a new app, called "phyphox", which is specifically designed for utilizing experiments in physics teaching. The app is free and designed to offer the same set of features on Android and iOS.
\end{abstract}

\noindent{\it physics, smartphone experiments, physics education, sensors, data acquisition, demonstration experiments\/}


\maketitle

\section{Introduction}

New smartphone-based experiments for physics teaching are proposed on a regular basis and use a wide range of integrated sensors that are accessed with many different apps \cite{vogt_experiments_2011, pendrill_acceleration_2011, kuhn_smartphones_2013, chevrier_teaching_2013, vieyra_turn_2015}. While these experiments promise to be motivating for the students and allow for a novel approach to give students measurement tools without additional costs, there are some obstacles that physics teachers need to overcome to use these experiments in practice. For most experiments one or both of the following problems prevent them from unfolding their full potential in class:
\begin{enumerate}
\item[1] \textbf{The smartphone itself is inaccessible as it is part of the experimental setup.}\\
The students have to perform the experiment blindly as the phone swings on a pendulum or is hidden from view in some apparatus. They only get to see their data at the end, which often consists of a big set of data points that appear disconnected from the experiment that has just been done. If the students were able to see the data generated in real-time, they could associate different stages of the experiment to different parts of the plot as it forms simultaneously. Similarly, the screen of the phone cannot show the recorded data to an audience, making it difficult to use smartphones in demonstration experiments in lectures an at schools.
\item[2] \textbf{The data is incomprehensible until analyzed on a computer.}\\
The students may see the data being collected during the experiment, but they need to export the data to a computer to do further data analysis, which then ranges from tasks like finding the time code of a recorded event to numerical integration in a spreadsheet software like Excel. Since these often are not simple tasks for students, this usually requires not only detailed instructions and more time than the hands-on activity, but poses an enormous extraneous cognitive load which might hinder the learning progress concerning the physical content. At the end of the day, the student is often more familiar with the software used to analyze the data than with the physics of the actual experiment that has been performed.
\end{enumerate}

Either of both problems can limit the motivation and comprehension for the students that could be gained from smartphone-based experimentation. In extreme cases, the experiment can only serve as an introduction to data analysis rather than as a means to discover physics phenomena. This drawback may also result in only hesitant application of smartphone-based experiments by physics teachers.

\section{The app phyphox}

In order to address both problems, we have created a new app for experiments with smartphone sensors. It addresses problem 1 by adding a simple to use remote access function that allows to remotely control and observe real-time experimental data from any second device. Problem 2 is addressed by including data analysis within the app. But instead of creating a black box which just generates a result, we implemented the data analysis customizable, so that each step of the analysis can be reviewed and modified by each app user, i.e. by the teacher. This promises a huge didactic potential by allowing to adjust how much data analysis should be done by the students themselves to the needs of each class or group of learners in general in a specific learning situation.

Besides these features, the app should be usable by as many students as possible. Therefore, we established that it has to be free of charge, free of advertisement and that it should have the same functionality and interface on Android and iOS. While there are some technical limitations to the extent to which Android and iOS devices can have the same functionality (mostly due to the availability and accessibility of certain sensors), it is imperative that a teacher should not have to address students with iPhones separately from students with Android devices.

We named the app \textit{phyphox} (as an acronym for \textit{physical phone experiments}) and released it on Google's Play Store and Apple's App Store in September 2016. To support users around the world, we created an accompanying website at \textit{http://phyphox.org}, which offers detailed instructions, demonstration videos and technical information in English and German. The app itself is currently being translated into additional languages by volunteers from all around the world.

\section{Remote access}

There is a range of examples for experiments that do not allow students to see their measurement result permanently during the experiment. A common one is the phone in a pendulum setup \cite{vogt_analyzing_2012-1,kuhn_analyzing_2012,briggle_analysis_2013}. Other interesting examples are most experiments that measure centripetal acceleration by rotating a phone \cite{vogt_analyzing_2013, monteiro_angular_2014, hochberg_spinning_2014} (for example in a salad spinner or on a record player) and experiments in which the phone is placed in a tire or a roll to measure its velocity \cite{puttharugsa_investigation_2016, wattanayotin_investigation_2017, yan_variation_2018}.

We will use the latter example to explain the problem and our solution to it. All that is needed for this experiment is a roll, for example a part of a paper tube as used to transport posters, a smartphone and some padding, which can also be some paper. The padding is used to center the phone in the tube and then the phone's rotation rate sensor (which is commonly referred to as a gyroscope on smartphones) is used to determine the angular velocity of the roll. Using its radius, the actual speed of the roll can be calculated. Typically, this can be used to demonstrate the linear acceleration of the roll on an inclined plane or, for more advanced courses, to discuss the roll's moment of inertia.

When using common data acquisition apps, the measurement often has to be started before placing the phone in the tube and can only be stopped after the phone has been taken out again. As a result, students see a big dataset with artifacts from handling the tube and need to export the data to a computer to calculate the speed. Most importantly, they have to take out their phones to see the data, which makes it impractical to actually experiment for example with different inclinations, to optimize the placement of the phone or to just repeat a failed run. This drawback can also not be avoided by the fact that the phyphox app features a handy countdown function to start an experiment.

\begin{figure}[htb]
  \centering
  \includegraphics[height=0.35\textwidth]{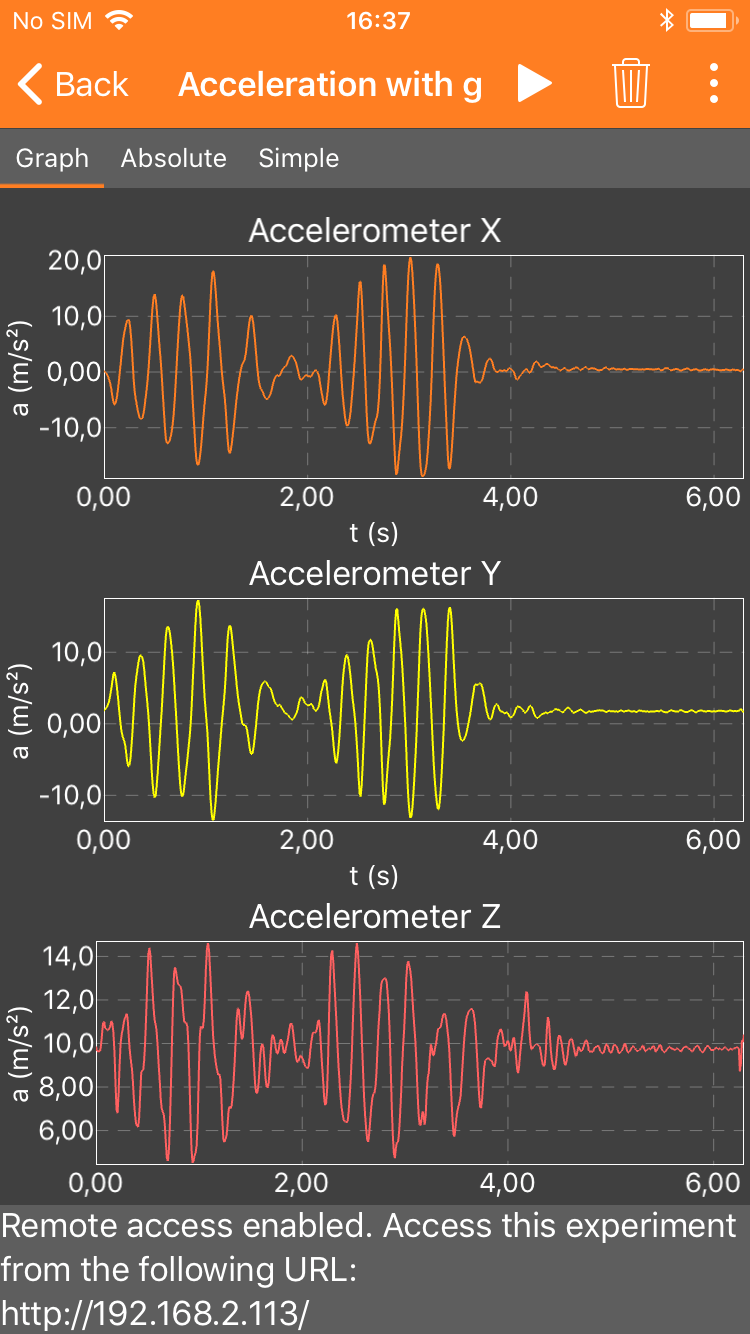}
  \includegraphics[height=0.35\textwidth]{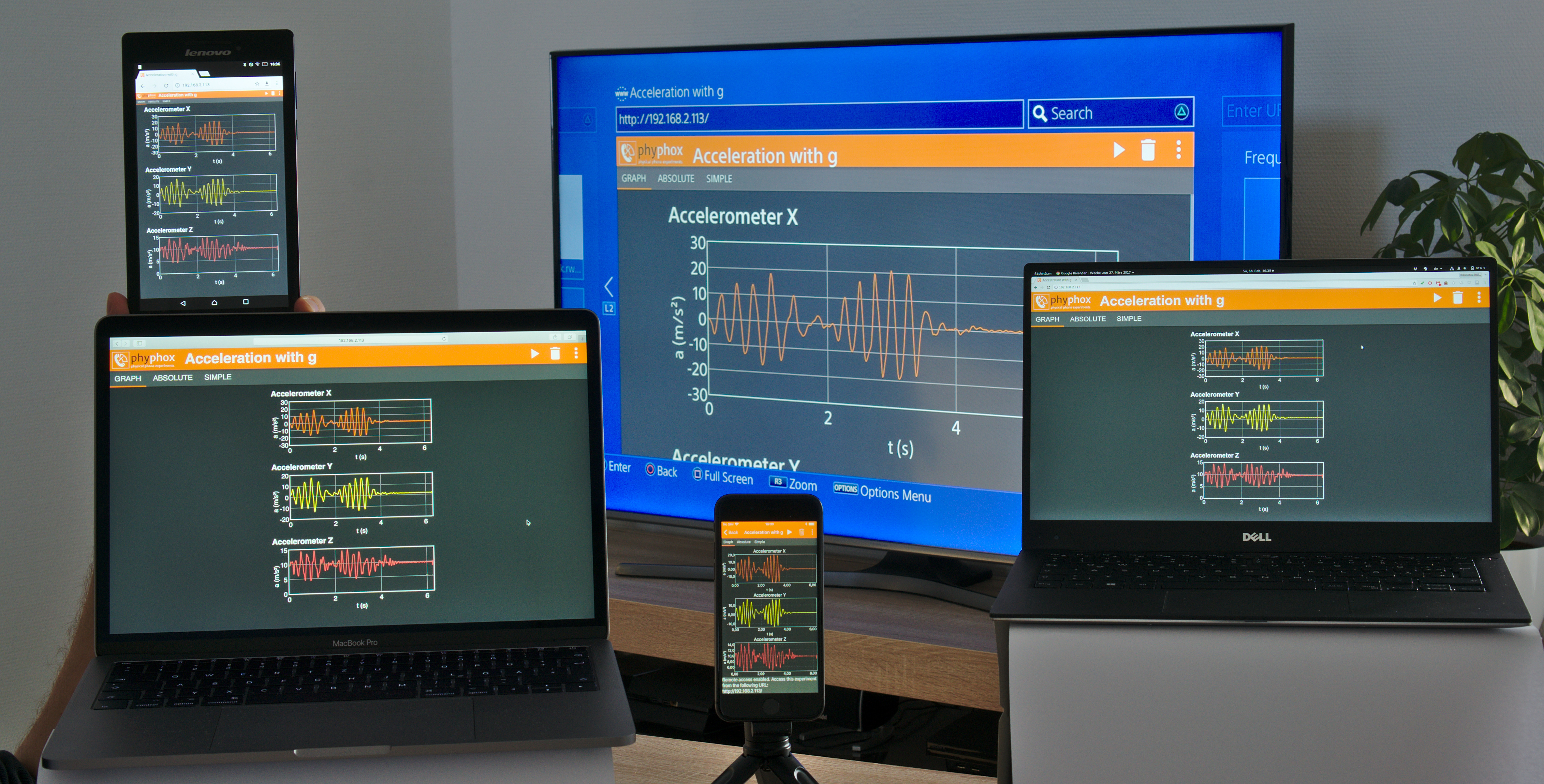}
	\caption{Remote access of phyphox. Left: Screenshot of phyphox with remote access feature enabled, showing an URL at the bottom of the screen. Right: The same device (an iPhone 8 in the center of the picture) is remotely accessed by an Android tablet (top left), a MacBook (bottom left), a Playstation 4 (background) and a laptop PC running GNU/Linux (right). The app only needs to be installed on the host device (in this case the iPhone).}
	\label{fig1}
\end{figure}

Instead, when doing this experiment in phyphox, the students select the experiment \textit{roll} within our app and then enable the function \textit{remote access} from the menu. The phone displays an URL (for example \textit{http://192.168.2.113} as shown in fig. \ref{fig1} (left)) which one enters into the address bar of a web browser on any other device which is in the same network (for example the phone of a second student, see fig. \ref{fig1} (right)). From then on, the second phone can start and stop the experiment and plot the collected data in real-time. Also, since phyphox can do data analysis as discussed later, the students can enter the radius of the roll into the app and get its speed in meter per second. They can start to experiment with the roll and extract its speed as their curiosity lets them explore the reactions of the graph on the second device.

Sometimes, this can also be achieved through video streaming. Any app can be remotely visible if the screen of the phone is streamed over the network, but there are several problems with this approach besides the high costs of apps which offer screen streaming. There are very few streaming apps which allow to control the phone from the remote side and video streaming causes high stress on the network. Most phones feature a high definition screen which needs to be encoded to a video stream and sent over the local wifi network. The result varies for different devices (iOS or Android, slow phone or a fast one with hardware accelerated video encoding, different screen resolutions) and always leads to a clearly noticeable latency between the experiment and the reaction on screen.

The phyphox approach is different as it does not stream a video, but just the pure experimental data. Technically, the app acts as a web server and the remote device just needs a modern web browser to connect to the app. Phyphox serves a regular web page which includes code that from then on polls the latest data from the app. On a good network connection the latency between experimental events and the reaction of the graph is barely noticeable and a simple wifi router can support many students at the same time. Additionally, the data representation fits the screen format of the target device, so the measured data can be shown to an audience via a laptop and a projector using the entire screen without being limited to a portrait screen format of the phone. Since the controlling device only needs a web browser, it can be anything from a desktop computer to another phone regardless of their operating system. On top of these benefits, the data can be directly downloaded from the web browser. If the students should do further data analysis, it can be downloaded immediately onto the computer on which they will do the next steps. The only requirement (which would be the same for most video streaming apps) is that both devices are on the same network.

\section{Data analysis}

In the example of the \textit{roll} experiment, we already mentioned the advantage of doing a simple mathematical operation to show the speed of the roll instead of showing its angular velocity. Again, there are several examples for smartphone-based experiments that need some data analysis which should not always be done by the students. A very common one that is often not recognized as data analysis performed by an app, is calculating the Fourier transform of an acoustical signal. A wide range of audio spectrum apps which accomplish this task are available and they are used in several smartphone experiments \cite{klein_classical_2014,hirth_measurement_2015,monteiro_measuring_2015}. Surprisingly, there are barely any apps that apply data analysis to other sensors than the microphone to simplify experimentation.

Applying data analysis in the app can not only make an experiment more accessible, it can also shift the focus of the experiment itself. An experiment that we showcase since the first release of phyphox tracks the movement of an elevator using the phone's atmospheric pressure sensor and its accelerometer. The atmospheric pressure recorded by the phone is used to calculate height differences, the (numerical) derivative of these height values yields a vertical speed and the accelerometer directly provides the vertical acceleration of the elevator.

When the students only record the raw acceleration and pressure data without in-app analysis, as proposed by Monteiro and Mart\'i \cite{monteiro_using_2017}, this is a great experiment to teach data analysis with the not too complicated example of applying the hydrostatic approximation to a large dataset and the optionally more advanced step of numerically deriving the result to obtain the velocity.

\begin{figure}[htb]
  \centering
  \includegraphics[width=0.3\textwidth]{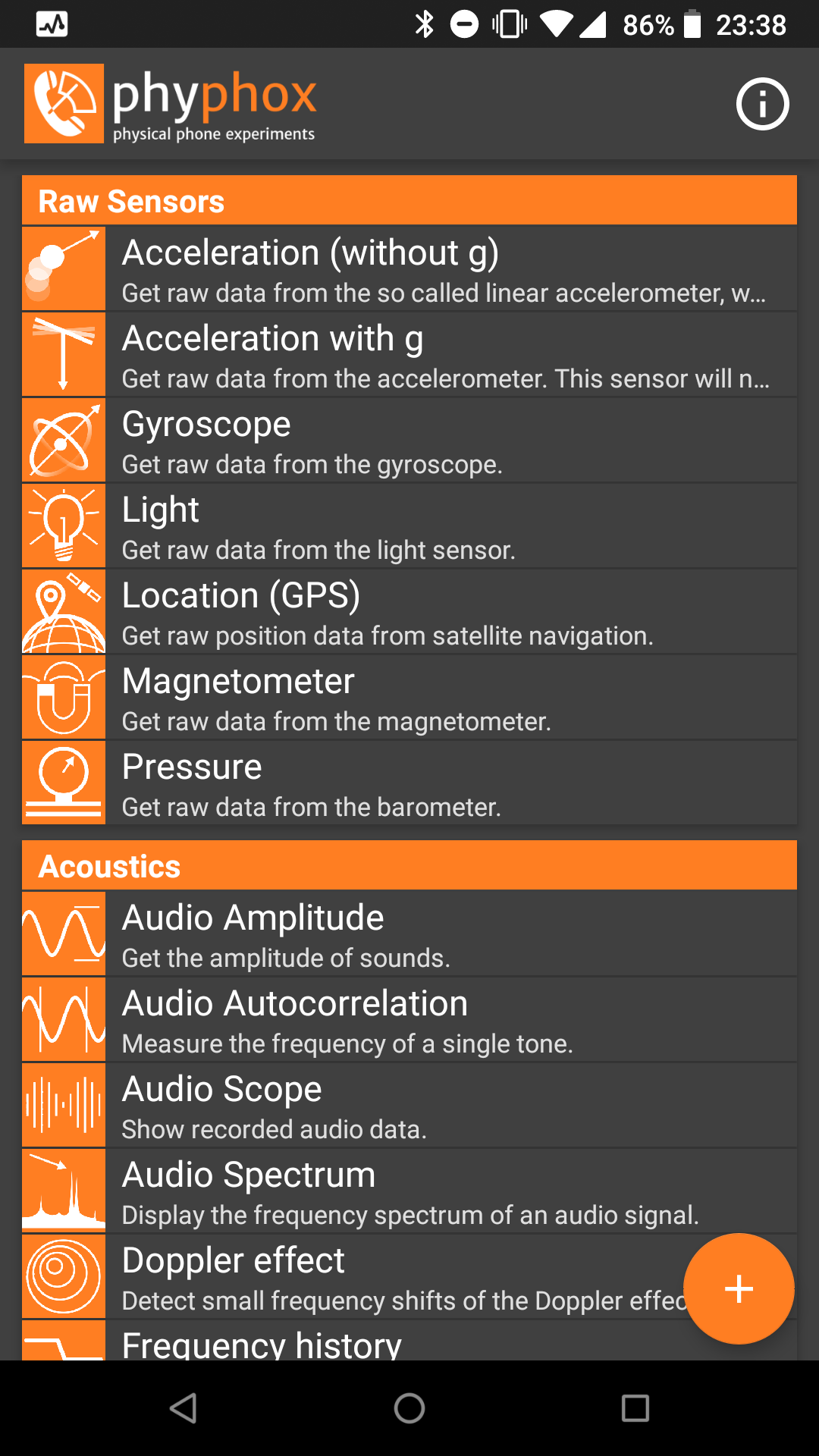}
  \includegraphics[width=0.3\textwidth]{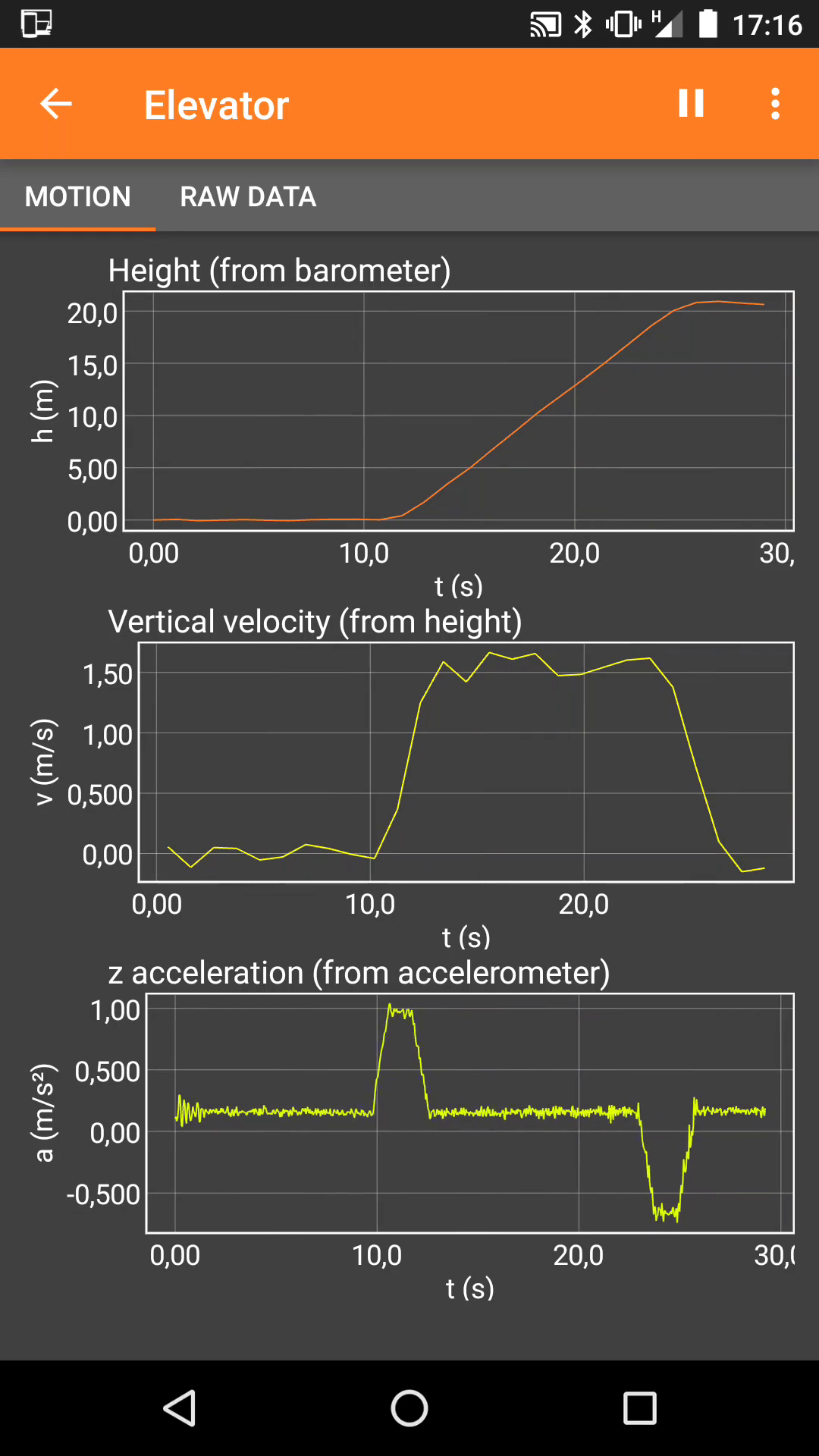}
  \includegraphics[width=0.3\textwidth]{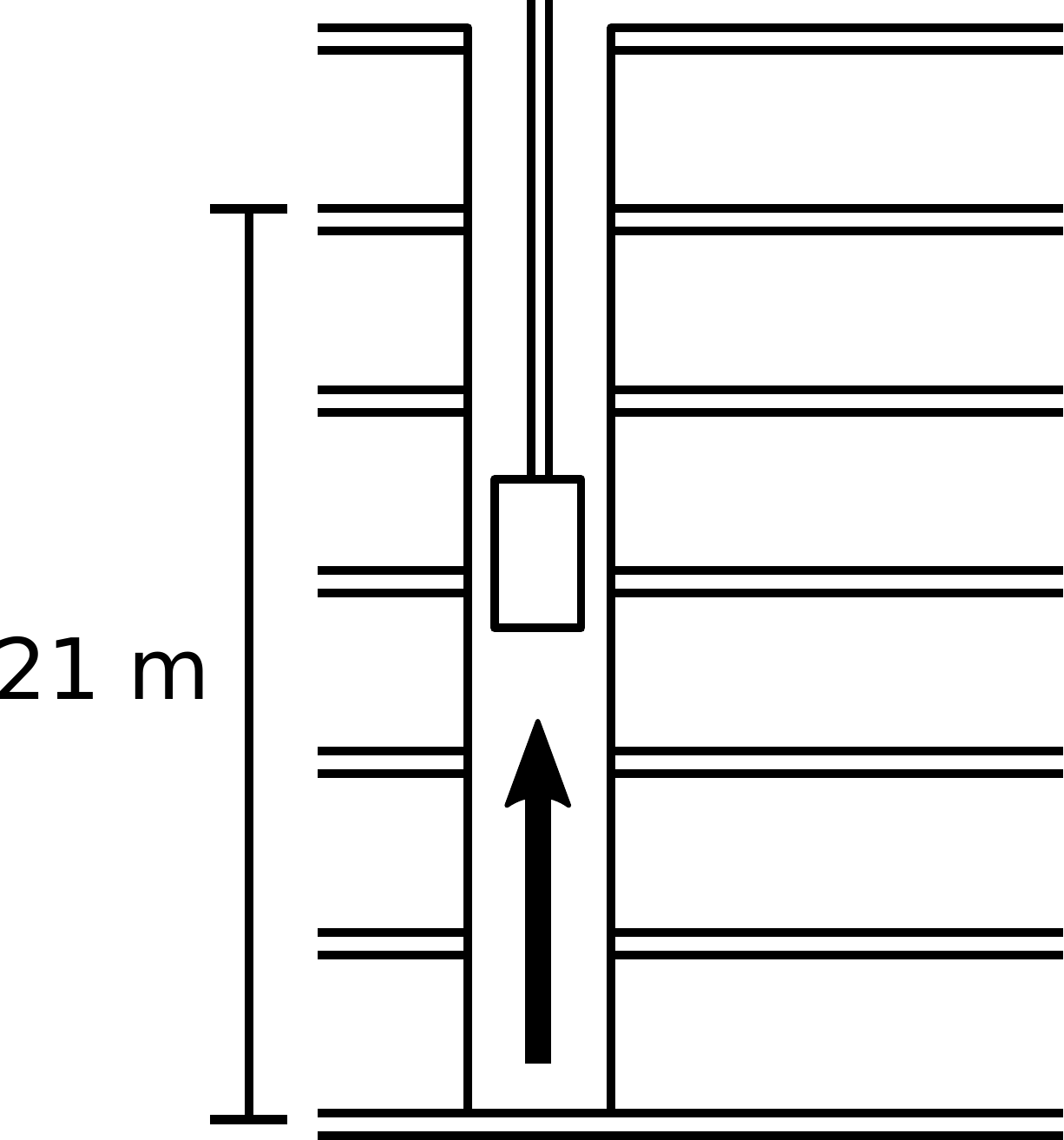}
	\caption{Screenshots of phyphox on Android. Left: The main menu offers access to the raw sensor data, followed by a big collection of experiments with data analysis, categorized by topics. Center: Result of the elevator experiment done in an elevator going up 5 floors. Right: Sketch of the elevator. Building height determined from the top floor with a laser rangefinder.}
	\label{fig2}
\end{figure}

In contrast, when using phyphox, all the analysis is done by the app and the students get graphs showing the altitude, vertical speed and acceleration as a function of time (see fig. \ref{fig2} (right)). In this case, the experiment becomes a tool to teach basic kinematics. The students easily understand the motion of an elevator and so they can understand these common plots very easily as they see in real-time how the changing altitude is plotted and how the acceleration they feel shows up immediately in the acceleration plot. This experiment does not require any additional knowledge, nor does it require access to computers to analyze the data.

Both approaches address different teaching situations and teach different skills. Automated data analysis within the app allows students to focus on different aspects of a physics experiment and allows to do some experiments with students who do not yet have the required mathematical skills. An extreme example is a version of the elevator experiment that we have created for a science museum. As their visitors include young children and people without any background in math or science, we have created a version, which just shows the height difference as a numerical value without any graphs. The experiment thus turns into a very accessible outreach outside of any classroom which sparks interest in how a height difference can be obtained.

Such customizations can easily be created by any user of phyphox. The app comes with a whole range of ready to use experiments (such as the \textit{roll} and \textit{elevator} experiment, see fig. \ref{fig2} (left)), but none of them are actually programmed as a fixed part of the app. Instead, phyphox uses its own file format to describe these experiments. A phyphox experiment file defines data sources (such as the accelerometer or the pressure sensor), sets up a range of mathematical operations on the data (from a simple addition to a complex Fourier transform) and displays the result by different means of visualization (like numerical values or graphs). The file format is XML-based and entirely documented on \textit{http://phyphox.org}.

\begin{figure}[htb]
  \centering
  \includegraphics[width=\textwidth]{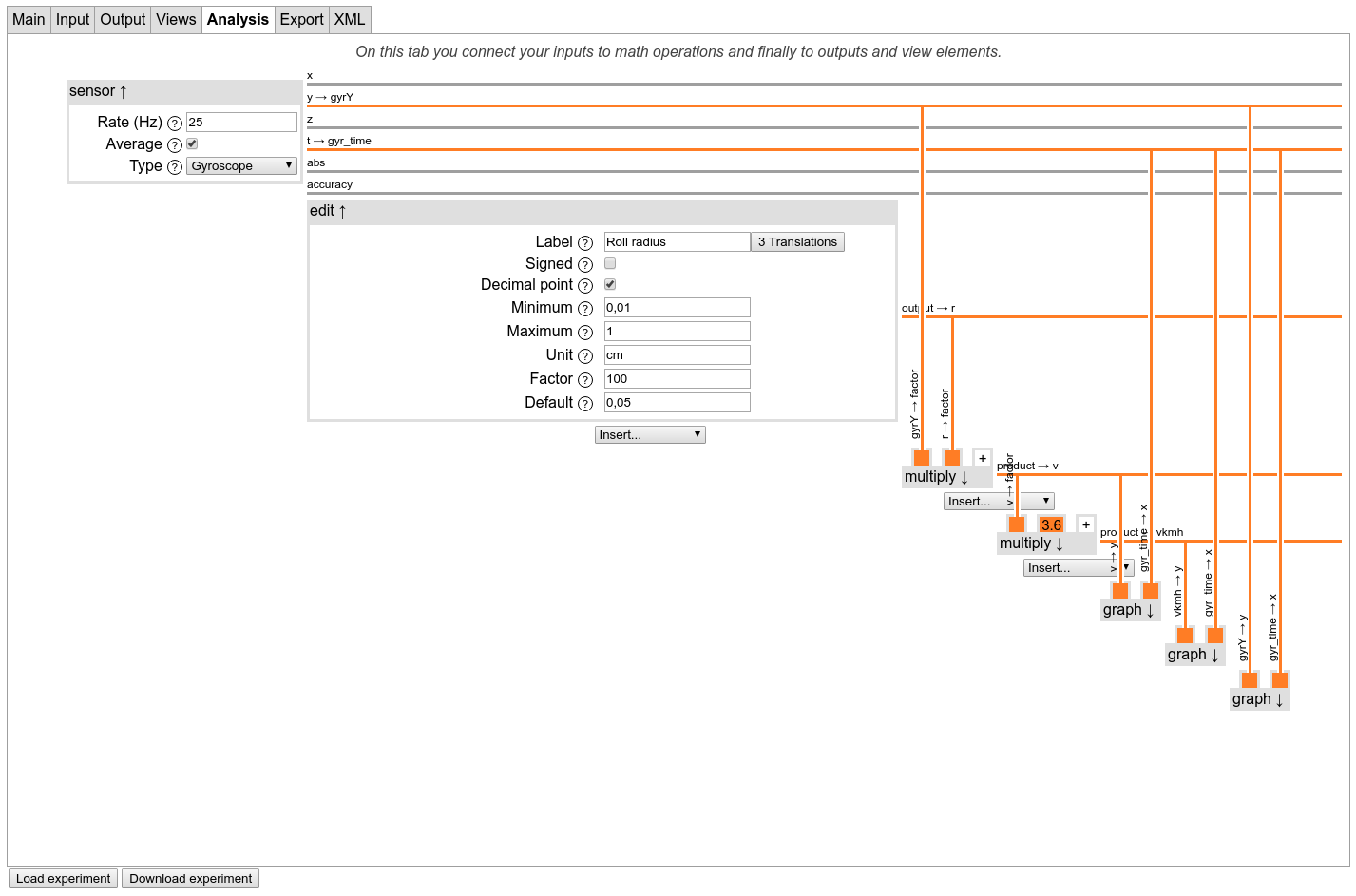}
	\caption{Screenshot of the web-based editor. The experiment \textit{roll} is shown, which uses the gyroscope's y axis and a radius entered by the user (into a so-called \textit{edit} field) to calculate and show the roll's speed.}
	\label{fig3}
\end{figure}

For a simple customization of existing experiments, there is also a web-based editor on our website (see fig. \ref{fig3}), which allows to open any of the existing experiments and to modify them using an intuitive visual interface. The resulting customized experiment can be downloaded onto a computer and then transferred to the phone through third party apps (email, messenger apps, cloud sharing, Bluetooth file transfer etc.) or distributed on a website. In this way, a teacher can customize smartphone assisted experiments to the state of knowledge and the educational objectives in the specific learning arrangement. This holds also for large groups of learners as for hundreds of students in an introductory physics course at university level.

\section{Conclusions}

Smartphone-based experiments can motivate students as it allows them to explore physics with their own tools. Our free app, phyphox, makes a variety of experiments more accessible and extends the tools available to students with a simple method to remote control the experiment and with data analysis in the field. While only reading raw data from the sensors can be great to teach data analysis, it often distracts from understanding the physical background of an experiment and sets up additional requirements to the mathematical skills of the students. With phyphox these requirements can either be removed entirely or tailored to the needs of a specific class and a specific learning situation.

\bibliographystyle{unsrt}

\bibliography{refs}{}

\end{document}